\documentstyle[12pt,epsf]{article}
\newcommand{\eps}{\epsilon}

\newcommand{\II}{{\cal I}}
\newcommand{\FF}{{\cal F}}

\newcommand{\GG}{{\cal G}}

\newcommand{\wt}{\widetilde}
\newcommand{\wh}{\widehat}

\newcommand{\be}{\begin{equation}}
\newcommand{\ee}{\end{equation}}
\newcommand{\ben}{\begin{eqnarray}\displaystyle}
\newcommand{\een}{\end{eqnarray}}
\newcommand{\refb}[1]{(\ref{#1})}
\newcommand{\p}{\partial}
\newcommand{\sectiono}[1]{\section{#1}\setcounter{equation}{0}}

\begin{document}

{}~ \hfill\vbox{\hbox{hep-th/9909062}\hbox{MRI-PHY/P990926}
}\break

\vskip 3.5cm

\centerline{\large \bf Supersymmetric World-volume Action for}
\medskip
\centerline{\large \bf Non-BPS D-branes}

\vspace*{6.0ex}

\centerline{\large \rm Ashoke Sen
\footnote{E-mail: asen@thwgs.cern.ch, sen@mri.ernet.in}}

\vspace*{1.5ex}

\centerline{\large \it Mehta Research Institute of Mathematics}
 \centerline{\large \it and Mathematical Physics}

\centerline{\large \it  Chhatnag Road, Jhoosi,
Allahabad 211019, INDIA}

\vspace*{4.5ex}

\centerline {\bf Abstract}

We construct the world-volume action for non-BPS D-branes in type II
string theories. This action is invariant under all the unbroken
supersymmetries in the bulk, but these symmetries are realised
as spontaneously broken symmetries in the world-volume theory. Coupling of
this action to background supergravity fields is straightforward.
We also
discuss the fate of the U(1) gauge field on the D-brane world-volume after
tachyon condensation.

\vfill \eject

\baselineskip=17.9pt

\tableofcontents

\sectiono{Introduction and Summary} \label{s1}

During the last year it has been realised that type IIA (IIB) string
theory admits unstable non-BPS D-branes of odd (even) 
dimensions\cite{9809111,9812031,9812135}.
These
D-branes can give rise to stable non-BPS D-branes after we mod out the
original type II string theory by some discrete symmetry 
group\cite{9805019,9806155,9808141,9809111,9810188,9812031,9901014}.
Since the
non-BPS D-brane is not invariant under any of the space-time
supersymmetry transformations, the spectrum on the world-volume of these
D-branes does not have Bose-Fermi degeneracy in general, although in some
special
cases the spectrum can develop accidental Bose-Fermi
degeneracy\cite{9908060}.

However, although there is no manifest supersymmetry in the world-volume
theory, we still  expect the world-volume theory to be supersymmetric,
with the supersymmetry realised as a spontaneously broken
symmetry.\footnote{I wish to thank O. Aharony, B. Kol and Y. Oz for
discussion on this
point.} The supersymmetry generators of the bulk theory, acting on the
non-BPS D-brane, produce fermion zero modes which can be identified as
the goldstinoes associated with the spontaneously broken supersymmetry.
Indeed, even for an ordinary BPS D-brane, which is invariant under half of
the space-time supersymmetry of the bulk theory\cite{9510017}, the
world-volume theory
possesses all the supersymmetries of the bulk theory, with half of them
realised as unbroken symmetries, and the other half realised as
spontaneously broken symmetries.

In this paper we shall construct the supersymmetric generalization of the
Dirac-Born-Infeld (DBI) action describing the dynamics of light modes on
the world-volume of a non-BPS D-brane. In section \ref{s2} we focus on the
world-volume theory of the massless fields of non-BPS D-brane of type II
string theory in Minkowski space-time.  As we shall see, this can be
constructed
with almost no work,
using the $\kappa$-symmetric action for a BPS D-brane derived in
ref.\cite{9512062,9610148,9610249,9611159,9611173,9612080}. In the
convention of refs.\cite{9611159,9611173,9612080} the $\kappa$ symmetric
action for a
BPS D-brane has two parts, $-$ the supersymmetric DBI action and the
Wess-Zumino term.  Each of these terms
is separately
invariant under the full set of space-time supersymmetry transformations,
but only the combined system has $\kappa$ gauge invariance.  To begin
with, the action has double the number of fermionic degrees of freedom
compared to the number of physical degrees of freedom on a BPS D-brane,
but fixing the $\kappa$ gauge symmetry removes half of these degrees of
freedom. We show that the world-volume action describing the non-BPS
D-brane is given by just the supersymmetric DBI part of the $\kappa$
symmetric action
describing a BPS D-brane. By construction the action is invariant under
all the space-time supersymmetries. But it does not have the $\kappa$
gauge symmetry, and hence the number of physical massless fermionic
degrees of
freedom is exactly double of that on a BPS D-brane. This is precisely
the case for a non-BPS D-brane. The open strings living on a non-BPS
D-brane have an extra sector in which the GSO projection is reversed, and
hence there are double the number of massless fermionic degrees of freedom
on a non-BPS D-brane compared to that on a BPS
D-brane\cite{9806155,9808141,9809111}.

If we want to construct the world-volume action of a non-BPS D-brane on an
orbifold or an orientifold of type II string theory, then typically the
action is obtained from the action described above by removing the degrees
of freedom which are projected out under the orientifold or the orbifold
operation. However, in some cases we may get extra light degrees of
freedom after the orbifold/orientifold operation. Thus for example a type
I 0-brane acquires extra fermionic zero modes from open string stretched
between the 0-brane and the 9-brane\cite{9808141,9809111}; these degrees
of freedom are
not present in the 0-brane of type IIB string theory. Another example
involves non-BPS D-brane of type II string theory on a K3 orbifold. In
this case at special points in the moduli space of K3 the world-volume
theory of
the brane may contain extra massless scalar fields which are not
present in the non-BPS D-branes of type II string theory in the ten
dimensional Minkowski space-time. In section \ref{s3} we discuss
inclusion of these fields in
the world-volume action maintaining supersymmetry of the world-volume
action.

In section \ref{s4} we discuss the effect of including the tachyon in
the world-volume action of the non-BPS D-brane of type II string theory in
Minkowski space-time. In this case the tachyon mass$^2$ is of the order of
the string scale, and hence there is no systematic way of constructing the
effective world-volume action involving the tachyon. But we can still
write down the general form of the action that maintains supersymmetry.
{}From general arguments one expects that at the minimum of
the tachyon potential the configuration is indistinguishible from the
vacuum\cite{9805170}, and hence the U(1) gauge field and the other
degrees of freedom on the D-brane
world-volume should disappear; but
exactly
how this happens has not been completely
understood\cite{9807138,9810188,9901159}. We show that if we ignore terms
involving derivatives of the gauge field strength and the acceleration of
the brane, then at the minimum of the tachyon potential the world-volume
action vanishes identically. Thus the gauge field now acts as a lagrange
multiplier field which imposes the constraint that the U(1) gauge current
must
vanish identically. This would explain the disappearance of the U(1) gauge
field at the tachyonic ground state, but in order to reach a definitive
conclusion we need to study the effect of the terms in the world-volume
action involving higher derivatives which were ignored in our analysis. 

\sectiono{World-volume action of massless fields for non-BPS
D-branes in type II string theory} \label{s2}

We shall begin with type IIA string theory, which admits BPS D-branes of
even dimension and non-BPS D-branes of odd dimension. Our focus of
attention will be a non-BPS D$p$-brane with $p$ odd, and we shall attempt 
to construct the world-volume action involving the massless fields on the 
brane by integrating out all the massive modes, including the 
tachyon.\footnote{Throughout this paper we shall be
working at the open string tree level, so integrating out the massive
modes amounts to eliminating them using their equations of motion.}
Throughout
this section we shall consider the background space-time to be ten
dimensional Minkowski space with no background fields, but as we shall
point out at the end of the section, coupling to background supergravity
fields is straightforward. Let
$\sigma^\mu$ ($0\le\mu\le p$) denote the coordinates on the world-volume
of the D-brane. 
The open strings living on such a D-brane has two Chan Paton (CP) sectors,
the sector labelled by the $2\times 2$ identity matrix $I$, and the sector
labeled by the Pauli matrix $\sigma_1$\cite{9809111}. The massless
dynamical
degrees of freedom on the D-brane world-volume are a set of 10 bosonic
coordinate fields $X^M(\sigma)$ $(0\le M\le 9)$, a U(1) gauge field
$A_\mu$, and a 32 component fermionic field $\theta$ which transforms as a
Majorana spinor under the space-time Lorentz group SO(9,1), but is a
world-volume scalar. The field $\theta$ can be regarded as the sum of a
left-handed Majorana-Weyl fermion $\theta_L$ and a right-handed
Majorana-Weyl fermion $\theta_R$. Of these all
the fields except $\theta_L$
come from the
CP sector $I$, and $\theta_L$ comes from the sector
$\sigma_1$.\footnote{Of course by changing our convention we could
have
the right-handed component of $\theta$ come from sector $\sigma_1$ and the
left-handed component come from the sector $I$.} For comparison let us
note that the spectrum
of massless fields coming from the sector $I$ is identical to that on
the world-volume of
a BPS D$p$ brane of type IIB string theory.
Thus for the BPS D$p$-brane the fermionic field transforms in the
Majorana-Weyl representation rather than in a Majorana representation.

The world-volume action of the non-BPS D-brane involving
these fields must satisfy the following criteria: 
\begin{enumerate}
\item It must be invariant under all the global supersymmetries of type
IIA string theory in the bulk. The supersymmetry transformation parameter
is a Majorana spinor $\eps$ of SO(9,1).
\item If we set $\theta_L=0$, then we are left with only the fields
originating in the identity sector. Except for an overall normalization
factor representing the difference in the tension of a non-BPS and a BPS
D-brane, the resulting effective action must agree with that on a BPS
D$p$-brane of type IIB string theory. This follows from the observation
that the rules for computing the open string amplitudes on a non-BPS
D-brane are identical to that on the BPS D-brane except for the presence
of the Chan-Paton factors\cite{9809111}, and a CP factor $I$ only
contributes an
overall
normalization factor to the amplitude.
\end{enumerate}

We shall now use these guidelines to construct the world-volume action on
the non-BPS D$p$-brane of type IIA string theory. 
Let us define:
\be \label{e1}
\Pi^M_\mu = \p_\mu X^M -\bar\theta \Gamma^M \p_\mu\theta\, ,
\ee
\be \label{e2}
\GG_{\mu\nu}=\eta_{MN}\Pi_\mu^M\Pi_\nu^N\, ,
\ee
and 
\be \label{e3}
\FF_{\mu\nu}=F_{\mu\nu} -[\bar\theta \Gamma_{11}\Gamma_M\p_\mu\theta
(\p_\nu X^M-{1\over 2}\bar\theta\Gamma^M\p_\nu\theta)
-(\mu\leftrightarrow\nu)]\, ,
\ee
where $\Gamma^M$ denote the ten dimensional gamma matrices, $\Gamma_{11}$
is the product of all the gamma matrices, $\eta_{MN}$ is the ten
dimensional Minkowski metric with signature $(-1,1,\ldots 1)$, and
\be \label{e4}
F_{\mu\nu}=\p_\mu A_\nu-\p_\nu A_\mu\, .
\ee
We now claim that the following action satisfies the two conditions listed
above:
\be \label{e5}
S = -C \int d^{p+1}\sigma \sqrt{-\det(\GG_{\mu\nu}+\FF_{\mu\nu})}\, ,
\ee
where $C$ is a constant equal to the D$p$-brane tension.

First let us check that it has the required amount of supersymmetry. For
this note that this action has the same structure as the first term of the
$\kappa$ symmetric action of a BPS D-brane of type IIA string theory as
discussed in ref.\cite{9612080}, except that $p$ is odd instead of even in
the
present case. As shown in \cite{9612080}, both $\GG_{\mu\nu}$ and
$\FF_{\mu\nu}$
are invariant under the supersymmetry transformations:
\ben \label{e6}
&& \delta_\eps\theta=\eps, \qquad \delta_\eps X^M=\bar\eps \Gamma^M\theta,
\nonumber \\
&& \delta_\eps A_\mu=\bar\eps\Gamma_{11}\Gamma_M\theta\p_\mu X^M
-{1\over
6}(\bar\eps\Gamma_{11}\Gamma_M\theta\bar\theta\Gamma^M\p_\mu\theta
+\bar\eps\Gamma_M\theta\bar\theta\Gamma_{11}\Gamma^M\p_\mu\theta)\, ,
\een
where the supersymmetry transformation parameter $\eps$ is a Majorana
spinor of SO(9,1) Lorentz group. Since $\GG_{\mu\nu}$ and $\FF_{\mu\nu}$
are invariant under the supersymmetry transformation, the action $S$
defined in \refb{e5} is also invariant under this transformation.

In order to check that this reproduces the world-volume action of the
BPS D$p$-brane of type IIB string theory when we set the
fermionic field $\theta_L$ coming from the CP sector $\sigma_1$ to zero,
let us define
$\theta_L$ and $\theta_R$ through the relations:
\be \label{e7}
\theta=\theta_L+\theta_R, \qquad
\bar\theta_L\Gamma_{11}=\bar\theta_L, \qquad 
\bar\theta_R\Gamma_{11}=-\bar\theta_R\, .
\ee
As discussed earlier, we can take $\theta_R$ to originate in the identity
sector and $\theta_L$ to originate in the $\sigma_1$ sector. Setting
$\theta_L$ to zero gives us
\be \label{e8} 
\GG_{\mu\nu}+\FF_{\mu\nu}=\eta_{MN}\p_\mu X^M\p_\nu X^N + F_{\mu\nu}
-2\bar\theta_R\Gamma_M\p_\mu X^M\p_\nu\theta_R
+(\bar\theta_R\Gamma^M\p_\mu\theta_R)(\bar\theta_R
\Gamma_M\p_\nu\theta_R)\, .
\ee
The action obtained by substituting this into eq.\refb{e5} agrees with the
world-volume action of a BPS D$p$-brane of type IIB string theory after
fixing the $\kappa$ gauge symmetry\cite{9612080}. The $\theta_R$ appearing
in
\refb{e8} has to be identified to the field $\lambda$ of
ref.\cite{9612080}.

Besides supersymmetry, the action \refb{e5} also has several other
symmetries. They include space-time translation symmetry,
\be \label{e9}
\delta_\xi X^M=\xi^M, \qquad \delta_\xi\theta=0, \qquad \delta_\xi
A_\mu=0\, ,
\ee
and the SO(9,1) Lorenz symmetry,
\be \label{e10}
\delta_\Lambda X^M=\Lambda^M_{~N}X^N, \qquad \delta_\Lambda \theta
=R(\Lambda)\theta, \qquad \delta_\Lambda A_\mu =0\, .
\ee
Here $\Lambda^M_{~N}$ denotes an infinitesimal element of the so(9,1)
algebra, and $R(\Lambda)$ is the Majorana spinor representation of
$\Lambda$.
Finally, the action is manifestly invariant under the reparametrization of
the world-volume coordinate $\sigma$:
\be \label{e11}
\sigma^\mu\to f^\mu(\sigma)\, ,
\ee 
for some set of functions $\{f^\mu(\sigma)\}$. $X^\mu$ and
$\theta$ transform as world-volume scalars and $A_\mu$
transforms as a world-volume vector under this transformation. \refb{e11}
represents a gauge
symmetry of the action. A convenient choice of gauge is the static gauge:
\be \label{e12}
\sigma^\mu = X^\mu \qquad\hbox{for}\qquad 0\le\mu\le p\, .
\ee
If we define
\be \label{e13}
\phi^i=X^i\qquad\hbox{for}\qquad p+1\le i\le 9\, ,
\ee
then $\GG_{\mu\nu}+\FF_{\mu\nu}$ can be rewritten in the static gauge as:
\ben \label{e14}
\GG_{\mu\nu}+\FF_{\mu\nu}&=&\eta_{\mu\nu}+\delta_{ij}\p_\mu \phi^i\p_\nu
\phi^j + F_{\mu\nu} \nonumber \\
&& -2\bar\theta_L(\Gamma_\nu+\Gamma_i\p_\nu \phi^i)\p_\mu\theta_L
-2\bar\theta_R(\Gamma_\mu+\Gamma_i\p_\mu \phi^i)\p_\nu\theta_R \nonumber
\\
&& +(\bar\theta_L\Gamma^M\p_\mu\theta_L)(\bar\theta_L
\Gamma_M\p_\nu\theta_L+\bar\theta_R
\Gamma_M\p_\nu\theta_R)\nonumber 
\\
&& +(\bar\theta_R\Gamma^M\p_\nu\theta_R)(\bar\theta_L
\Gamma_M\p_\mu\theta_L+\bar\theta_R
\Gamma_M\p_\mu\theta_R)\, .
\een
If we set $\theta_L=0$, and rename $\theta_R$ as $\lambda$, the action
agrees with the action of BPS D$p$-brane of type IIB string theory in the
static gauge, as given in ref.\cite{9612080}.

The gauge fixed action is no longer invariant under the symmetry
transformations \refb{e6}, \refb{e9}, \refb{e10}. Instead, each of these
symmetry transformations must be accompanied by a compensating gauge
transformation which brings us back to the gauge $X^\mu=\sigma^\mu$. Thus
we must define the new symmetry transformations $\wh\delta$ as a
combination of the old transformations $\delta$ and a world volume
reparametrization, such that
\be \label{e15}
\wh\delta X^\mu(\sigma) = 0\, .
\ee
Thus the new transformation laws $\wh\delta$ for any
world-volume scalar field
$\Phi$ ({\it e.g.} $X^M$ or $\theta$) will be given in terms of the old
transformation laws $\delta$ as
follows:
\be \label{e16}
\wh\delta_{\eps,\xi,\Lambda} \Phi =\delta_{\eps,\xi,\Lambda}\Phi -
(\Delta_{\eps,\xi,\Lambda}^\mu) \p_\mu \Phi\, ,
\ee
where
\be \label{e17}
\Delta_\eps^\mu =
\bar\eps
\Gamma^\mu\theta, \qquad
\Delta_\xi^\mu = \xi^\mu , \qquad
\Delta_\Lambda^\mu =
(\Lambda^\mu_{~\nu}\sigma^\nu+\Lambda^\mu_{~i}\phi^i)\, .
\ee
If we take $\Phi=X^\mu$, then
it is easy to check using eqs.\refb{e6}, \refb{e9}, \refb{e10}, \refb{e16}
and \refb{e17} that eq.\refb{e15} is indeed satisfied.
For world-volume vector field $A_\mu$ the right hand side of eq.\refb{e16}
contains an
extra term $-(\p_\mu\Delta^\rho_{\eps,\xi,\Lambda})A_\rho$.

Let us now turn to non-BPS D$p$-branes of type IIB string theory. In this
case $p$ is even. The bosonic massless fields again consist of the
scalar fields $X^M(\sigma)$ and the gauge fields $A_\mu(\sigma)$,
but now, instead of a fermionic field
$\theta$ transforming as a Majorana spinor of SO(9,1), we have a pair of
fermionic fields $\theta_1$ and $\theta_2$, each transforming in
the right-handed Majorana-Weyl spinor representation of
SO(9,1).\footnote{Since the fermion zero modes from CP factor $\sigma_1$
carry opposite GSO projection compared to those from CP factor $I$, one
might wonder how we can get a pair of fermion fields of same SO(9,1)
chirality. For this note that in the static gauge (in which the open
string spectrum is computed) only an SO($p$,1)$\times$SO($9-p$) subgroup
of the full SO(9,1) Lorentz group is realized as a manifest symmetry.
Since $p$ is even for type IIB string theory, neither SO($9-p$) nor
SO($p,1$) has chiral spinor representation. As a result,
both a left-handed and a
right-handed Majorana-Weyl spinor of SO(9,1) will transform in the same
representation of SO($p,1$)$\times$SO($9-p$). Thus knowing the GSO
projection rules we cannot determine the SO(9,1) chirality of the fermion
fields. It is determined by requiring that these fermions represent the
goldstino fields associated with spontaneously broken supersymmetry.}
We define 
\be \label{e18} \theta=\pmatrix{\theta_1\cr\theta_2}\, , 
\ee
and let $\tau_3$ denote the matrix $\pmatrix{I & \cr & -I}$ acting on
$\theta$, where $I$ denotes the identity matrix acting on $\theta_1$ and
$\theta_2$. We also define 
\be \label{e1a}
\Pi^M_\mu = \p_\mu X^M -\bar\theta \Gamma^M \p_\mu\theta\, ,
\ee
\be \label{e2a}
\GG_{\mu\nu}=\eta_{MN}\Pi_\mu^M\Pi_\nu^N\, ,
\ee
\be \label{e3a}
\FF_{\mu\nu}=F_{\mu\nu} -[\bar\theta \tau_3\Gamma_M\p_\mu\theta
(\p_\nu X^M-{1\over 2}\bar\theta\Gamma^M\p_\nu\theta)
-(\mu\leftrightarrow\nu)]\, ,
\ee
\be \label{e5a}
S = -C \int d^{p+1}\sigma \sqrt{-\det(\GG_{\mu\nu}+\FF_{\mu\nu})}\, ,
\ee
where $C$ is a constant equal to the D$p$-brane tension.
It is easy to check that $\GG_{\mu\nu}$ and $\FF_{\mu\nu}$, and hence $S$,
is invariant under the supersymmetry
transformation:
\ben \label{e6a}
&& \delta_\eps\theta=\eps, \qquad \delta_\eps X^M=\bar\eps \Gamma^M\theta,
\nonumber \\
&& \delta_\eps A_\mu=\bar\eps\tau_3\Gamma_M\theta\p_\mu X^M
-{1\over
6}(\bar\eps\tau_3\Gamma_M\theta\bar\theta\Gamma^M\p_\mu\theta
+\bar\eps\Gamma_M\theta\bar\theta\tau_3\Gamma^M\p_\mu\theta)\, ,
\een
where the supersymmetry transformation parameter $\eps$ is given by
$\pmatrix{\eps_1\cr \eps_2}$, with $\eps_1$ and $\eps_2$ both right-handed 
Majorana
spinor of SO(9,1) Lorentz group. 

If we set $\theta_1=0$ and identify $\theta_2$ with $\lambda$, we recover
the world-volume action of a BPS
D$p$-brane of type IIA string theory after $\kappa$ gauge fixing, as
given in ref.\cite{9612080}. Thus the
action \refb{e5a} satisfies the required consistency conditions. As in
the case of type IIA string theory, this action also has
space-time translation
symmetry and ten dimensional Lorentz invariance; these transformation laws
are given by equations identical to \refb{e9}, \refb{e10}. Finally it has
world-volume reparametrization invariance, and using this we can go to a
static gauge $X^\mu=\sigma^\mu$. The gauge fixed action has a form
identical to that for type IIA non-BPS D-branes except that $\theta_L$ and
$\theta_R$ in eq.\refb{e14} are replaced by $\theta_1$ and $\theta_2$
respectively. The various transformation laws get modified in the static
gauge according to eqs.\refb{e16}, \refb{e17} as before.

We conclude this section by noting that the coupling of the
D-brane
world-volume theory discussed here to background supergravity fields can
be carried out following the procedure given in
ref.\cite{9611159,9611173}. Again for type IIA (IIB) string theory the
action is obtained by keeping
only the supersymmetric DBI term of the $\kappa$-symmetric action
describing BPS D-brane of type IIA (IIB) string theory in a supergravity
background, and taking $p$ to be odd (even) instead of even (odd).

\sectiono{Inclusion of other light fields} \label{s3}

The non-BPS D$p$-branes of type II string theory are unstable due to the
existence of a tachyonic mode on their world-volume\cite{9809111,9812031}.
But quite
often we can get stable non-BPS branes by taking certain
orientifolds/orbifolds of type II string theory, if this operation 
projects out the
tachyonic mode\cite{9806155,9808141,9809111,9812031,9901014}. Typically,
this will also project out a subset of
the massless degrees of freedom on the D-brane world-volume, and the
world-volume action of the resulting D-brane will be given by an
appropriate truncation
of the actions \refb{e5} or \refb{e5a}. But in some cases there are extra
(nearly) massless degrees of freedom on the world-volume, which, if
present, must be included in the world
volume action. Thus we need to know how to couple these fields maintaining
the ten dimensional super-Poincare invariance and world-volume
reparametrization invariance. We shall discuss two examples.

\subsection{Light scalars on non-BPS D-brane on $T^4/Z_2$ near a critical
radius} \label{s31}

This system was discussed in detail in
refs.\cite{9812031,9901014,9906109,9908060}.
We begin
with a
non-BPS D-brane of type IIA/IIB string theory on $T^4$ with an odd number
$n$
of tangential directions of the D-brane along $T^4$, and mod out the
resulting configuration by the $Z_2$ transformation $\II_4$ that reverses
the sign
of all four coordinates of the torus. At a generic point in the moduli
space of the torus the massless degrees of freedom consist of a subset of
the fields living on the D-brane before the orbifold projection. 
In order to simplify the action, we can do a partial gauge fixing by
identifying $n$
of the world-volume coordinates of the brane to the $n$ coordinates of the
torus along
which the brane extends. In this case these $n$ world-volume coordinates
will be compact, and we can dimensionally reduce the original world-volume
action by ignoring the dependence of all fields on these compact
coordinates. This gives
a $(p-n+1)$-dimensional world-volume theory. This dimensional
reduction has been
carried out in detail in \cite{9612080}. In this dimensionally reduced
action we then set to zero the fields which are odd under $\II_4$. We
shall continue
to denote by $\sigma^\mu$ the coordinates of this world-volume theory,
although
now $\mu$ runs over $(p-n+1)$ values.

An alternative procedure will be
to
make a series of $n$ T-duality transformations which converts the
D$p$-brane to a D$(p-n)$-brane, with all the $(p-n)$
tangential directions
of the brane lying along the non-compact
directions\cite{9812031,9901014}. Since $n$ is odd, this T-duality
transforms
$\II_4$ into $\II_4\cdot (-1)^{F_L}$ where $(-1)^{F_L}$ denotes the
contribution to the space-time fermion number from the left-moving sector 
of the closed string world-sheet\cite{9604070}. The world-volume action of
the brane is then obtained by starting from the world-volume action of the
non-BPS D$(p-n)$-brane discussed in the previous section, and setting to
zero all the fields which are odd under $(-1)^{F_L}\cdot \II_4$. Both
these procedures lead to the same result. 

At
certain critical values of the radii of the torus we can get one or more
extra massless scalar fields\cite{9812031,9901014,9906109}.\footnote{For 
a closely related
example in type I string theory, 
see refs.\cite{9808141,9902160,9903123}.} For simplicity
we
shall take only one of
the radii to be near the critical radius, so that we have only one nearly
massless scalar field. Let us denote this field by $\chi$. 
When all the other massless fields are set to zero, the low energy
effective action for $\chi$ takes the following form in the static gauge:
\be \label{e3p0}
{1\over 2}\int d^{p-n+1}\sigma (-\eta^{\mu\nu}\p_\mu\chi\p_\nu\chi
-V(\chi))\, ,
\ee
where
$V(\chi)$ is the tachyon potential constructed in \cite{9906109}.
In writing \refb{e3p0} we have ignored terms with more than two
derivatives of $\chi$.

We propose the
following supersymmetric and reparametrization invariant coupling of the
action \refb{e3p0} to other massless fields on the D-brane world-volume:
\be \label{ee1}
\int d^{p-n+1}\sigma \sqrt{-\det(\GG+\FF)} \, (- \wt 
\GG_S^{\mu\nu}\p_\mu\chi
\p_\nu\chi - V(\chi))\, ,
\ee
where $\wt\GG^{\mu\nu}$ denotes the matrix inverse of $(\GG+\FF)$, and
$\wt\GG_S$ denotes the symmetric part of $\wt\GG$. In writing \refb{ee1}
we have further ignored terms involving derivatives of $\GG$ and $\FF$. 
This action clearly satisfies the requirement of space-time
supersymmetry and world-volume reparametrization invariance provided we
take $\chi$ to be a world-volume scalar and inert under supersymmetry
transformation:
\be \label{ee2}
\delta_\eps\chi =0\, .
\ee
However, the requirement of supersymmetry alone does not fix the form of
the action. For example, instead of using the metric $\wt \GG_S^{\mu\nu}$,
we could have
used the metric $\GG^{\mu\nu}$, the matrix inverse of $\GG$, in the
term involving $\p_\mu\chi\p_\nu\chi$. In order to resolve this
ambiguity, we have used the result of ref.\cite{9908142}. 
Ref.\cite{9908142} analysed open string theory in the
presence of constant background metric and anti-symmetric tensor field,
and showed that the 
natural metric for open strings is the symmetric part of the inverse of
$(G+B)$ where $G_{\mu\nu}$ and $B_{\mu\nu}$ are respectively 
the pullback of the metric and the
antisymmetric tensor field on the D-brane world-volume. Since
in the D-brane world-volume action the gauge field always appears in the
combination
$(B_{\mu\nu}+F_{\mu \nu})$, we can use the
result of ref.\cite{9908142} to
conclude that in the presence of a constant background gauge field
strength
$F_{\mu\nu}$, the natural metric appearing in the kinetic term of $\chi$ 
is
the symmetric part of the inverse of $(G+F)$. The requirement of
space-time supersymmetry then fixes the form of the action
\refb{ee1}.\footnote{Of course the antisymmetric part of constant
background $\wt\GG^{\mu\nu}$ has the effect of making the ordinary
products in \refb{ee1} into non-commutative $*$
products\cite{9711162,9711165,9903205,9908142}.
This can be reexpressed in terms of ordinary products by including terms
with higher derivatives. In the present case these extra terms vanish due
to symmetry reasons.}

Note that although $\chi$ does not transform under supersymmetry in the
gauge invariant description, it does transform under supersymmetry
according to eqs.\refb{e16}, \refb{e17} in the static gauge.

\subsection{Fermionic zero modes on the type I D-particle} \label{s32}

Type I D-particle is obtained by modding out the type IIB D-particle by
the world-sheet parity transformation $\Omega$\cite{9808141,9809111}. The
massless
degrees of freedom on the world volume contains a subset of the
world-volume degrees of freedom of the type IIB D-particle which are
invariant under $\Omega$, and the effective action involving these modes
is given by the effective action of the type IIB D-particle with the
$\Omega$ odd modes set to zero. But type I D-particle also has 32 extra
fermionic zero modes from open strings stretched between the D0-brane and
the space filling D9-branes which are present in the type I string theory.
Let us denote these modes by $\psi^I$ ($0\le I\le 32$), and let us denote
the world-volume time coordinate of the D0-brane by $\tau$. We need to
construct a reparametrization and supersymmetry invariant world-volume
action for these modes. We propose the following action:
\be \label{ee3}
\int d\tau \psi^I\p_\tau\psi^I\, .
\ee
This is manifestly supersymmetry and reparametrization invariant if we
take $\psi^I$ to be a scalar under reparametrization, and inert under
supersymmetry transformation:
\be \label{ee4}
\delta_\eps \psi^I=0\, .
\ee
As in the previous case, $\psi^I$ acquires a non-trivial supersymmetry
transformation law in the static gauge according to the rules given in
eqs.\refb{e16}, \refb{e17}.

\sectiono{Effect of tachyon condensation on the non-BPS
D-brane of
type II string theory} \label{s4}

The non-BPS D-brane of type II string theory in ten dimensional Minkowski
space contains a tachyonic mode besides the massless modes and the
infinite number of massive modes. Let us consider the effective action
as a function of the tachyonic and the massless modes,
obtained by integrating out all the massive modes. Since
the tachyon mass$^2$ is of the order of the string tension, there is no
systematic procedure for computing this effective action, but we shall be
interested in studying some of the general  properties of this action. If
we ignore terms involving derivatives of $\GG_{\mu\nu}$ and
$\FF_{\mu\nu}$, then combining the results of \cite{9908142} with the
requirement
of supersymmetry, we can expect the following form of the effective
action:
\be \label{ee5}
S = -\int d^{p+1}\sigma \sqrt{-\det(\GG+\FF)} \, F(T, \p_\mu T,
D_\mu\p_\nu
T,
\ldots, \wt\GG^{\mu\nu}_S, \wt\GG^{\mu\nu}_A) + I_{WZ}\, .
\ee
Here $T$ is the tachyon field, and $F$ is some function of its arguments.
We have included the original action
\refb{e5}, \refb{e5a} into the definition of $F$ so that for $T=0$, $F=C$.
$\wt\GG^{\mu\nu}_S$ and $\wt\GG^{\mu\nu}_A$ denote the symmetric and the
antisymmetric
parts of
$\wt\GG\equiv(\GG+\FF)^{-1}$. The analysis of \cite{9908142} tells us that
the
background value of $\wt\GG_S$ gives the
natural
metric for the open string, whereas the effect of a background $\wt\GG_A$
is
to
convert an
ordinary product to a non-commutative $*$ product. $I_{WZ}$ denotes a
Wess-Zumino term representing the
supersymmetric generalization of
the coupling of the tachyon to
background Ramond-Ramond
fields\cite{9810188,9812031,9812135,9904207,9905157,9908029}.
This term typically involves the wedge product of $dT$ with a
supersymmetry invariant $p$-form on the D-brane world-volume. An
example of such a term will be the wedge
product of
$dT$ with the Wess-Zumino term for the BPS D$(p-1)$ brane constructed in
refs.\cite{9611159,9611173,9612080}.

We shall be interested in analysing this action for constant $T$. Thus
$\p_\mu T$ and hence $I_{WZ}$ vanishes. For such a background the
dependence of $F$ on
$\wt\GG_S$ and
$\wt\GG_A$ disappears, since there are no indices with which
$\wt\GG_S$ can
contract, and since for constant functions the $*$ product reduces to the
ordinary product. Thus the action can be rewritten as:
\be \label{ee6}
-\int d^{p+1}\sigma \sqrt{-\det(\GG+\FF)} \, V(T)\, ,
\ee
where $V$ is the tachyon potential. It has been argued on general grounds
that at the minimum $T_0$ of the potential $V$ vanishes, {\it i.e.}
\be \label{ee7}
V(T_0)=0\, .
\ee
{}From eq.\refb{ee6} we see that at $T=T_0$ the world volume action
vanishes identically. Thus in this case the world-volume gauge field acts
as a lagrange multiplier field which imposes the constraint that the U(1)
gauge current must vanish identically. As a result all states which are
charged under the U(1) ({\it e.g.} an open string with one end on this
non-BPS D-brane and the other end on another D-brane) will disappear from
the spectrum. 

This result is similar to the result of \cite{9901159} where it was argued
that
this U(1) gauge field is confined at the tachyonic
ground state.\footnote{Actually \cite{9901159} did
not look
at this problem, but to a closely related problem of tachyon condensation
on a brane-antibrane pair.} But the mechanism proposed in \cite{9901159}
was
non-perturbative from the point of view of the D-brane world-volume
theory, whereas the mechanism discussed here is a tree level effect in the
open string theory.

Since \refb{ee6} is the key result leading to the conclusion above, let us
review the origin of this equation in some detail. The main reason behind
this form of the effective action for constant $T$ is that inside the
function $F$ appearing in eq.\refb{ee5}, the indices of
$\wt\GG^{\mu\nu}_{S,A}$ always contract with
the derivative factors and not with each other. This follows from the
structure of open string disk amplitude. The effect of constant background
$\wt\GG$ is to modify the $X^\mu$ propagator on the boundary of the
disk. Since the $X^\mu$ dependence of the tachyon vertex operator comes
only from the momentum factors $e^{ik.X}$, $\wt\GG^{\mu\nu}$ can
modify only the momentum dependent factors of the correlation
function of tachyon
vertex operators. Thus for constant $T$ there is no non-trivial dependence
of the effective action on $\wt\GG^{\mu\nu}$.\footnote{This can also be
argued
using string field theory.}  The only dependence
on $\wt\GG^{\mu\nu}$ comes through the effective coupling constant 
of the open string theory\cite{9908142}, as an overall
multiplicative factor of $\sqrt{\det(\GG+\FF)}$. Since this argument works
only for
constant background $\wt\GG^{\mu\nu}$, it does not give us any
information about terms involving derivatives of $\wt\GG^{\mu\nu}$. 
Whether inclusion of these terms in our analysis
changes the conclusion remains to be seen. Unfortunately there does not
seem to be any systematic procedure for analysing these terms.

\noindent{\bf Acknowledgement}: I wish to thank O. Aharony, B.~Kol,
Y.~Oz and B. Zwiebach for
discussions.

\end{document}